\documentclass[a4paper,fleqn,english]{article} 
\usepackage{latexsym}\usepackage{graphicx}\usepackage{babel}
\usepackage{epsfig}
\begin{document}
\catcode`\@=11
\@addtoreset{equation}{section}
\def\theequation{\arabic{section}.\arabic{equation}}
\def\appendix{\renewcommand{\thesection}{\Alph{section}}\setcounter{section}{0}
 \renewcommand{\theequation}
 {\mbox{\Alph{section}.\arabic{equation}}}\setcounter{equation}{0}}
\def\schw{Schwarzschild }
\def\maketitle{\thispagestyle{empty}\setcounter{page}0\newpage
 \renewcommand{\thefootnote}{\arabic{footnote}}
 \setcounter{footnote}0}
\renewcommand{\thanks}[1]{\renewcommand{\thefootnote}{\fnsymbol{footnote}}
 \footnote{#1}\renewcommand{\thefootnote}{\arabic{footnote}}}
\newcommand{\preprint}[1]{\hfill{\sl preprint - #1}\par\bigskip\par\rm}
\renewcommand{\title}[1]{\begin{center}\Large\bf #1\end{center}\rm\par\bigskip}
\renewcommand{\author}[1]{\begin{center}\Large #1\end{center}}
\newcommand{\address}[1]{\begin{center}\large #1\end{center}}
\newcommand{\pacs}[1]{\smallskip\noindent{\sl PACS numbers:
 \hspace{0.3cm}#1}\par\bigskip\rm}
\def\babs{\hrule\par\begin{description}\item{Abstract: }\it} 
\def\eabs{\par\end{description}\hrule\par\medskip\rm}
\renewcommand{\date}[1]{\par\bigskip\par\sl\hfill #1\par\medskip\par\rm}
\newcommand{\ack}[1]{\par\section*{Acknowledgments} #1} 

\def\dinfn{Dipartimento di Fisica, Universit\`a di Trento\\ 
 and Istituto Nazionale di Fisica Nucleare,\\
 Gruppo Collegato di Trento, Italia \medskip}

\def\csic{Consejo Superior de Investigaciones Cient\'{\i}ficas \\
Instituto de Ciencias del Espacio (ICE/CSIC)\\
 Campus UAB, Facultat Ciencies, Torre C5-Parell-2a planta, \\
08193 Bellaterra (Barcelona), Spain \medskip}

\def\ieec{Institut d'Estudis Espacials de Catalunya (IEEC), \\
Edifici Nexus, Gran Capit\`a 2-4, 08034 Barcelona, Spain \medskip}

\def\icrea{$^d$ Instituci\`o Catalana de Recerca i Estudis Avan\c{c}ats 
(ICREA), Barcelona, Spain\medskip}

\def\guido{Guido Cognola\thanks{e-mail: \sl cognola@science.unitn.it\rm}}
\def\sergio{Sergio Zerbini\thanks{e-mail: \sl zerbini@science.unitn.it\rm}}
\def\luciano{Luciano Vanzo\thanks{e-mail: \sl vanzo@science.unitn.it\rm}}
\def\emilio{Emilio Elizalde\thanks{e-mail: \sl elizalde@ieec.uab.es\rm}}
\def\sergei{Sergei D.~Odintsov\thanks{e-mail: \sl odintsov@ieec.fcr.es\rm}}


\newcommand{\s}[1]{\section{#1}}
\renewcommand{\ss}[1]{\subsection{#1}}
\newcommand{\sss}[1]{\subsubsection{#1}}
\def\M{{\cal M}} 
\newcommand{\ca}[1]{{\cal #1}} 
\def\segue{\qquad\Longrightarrow\qquad} 
\def\prece{\qquad\Longleftarrow\qquad} 
\def\hs{\qquad} 
\def\nn{\nonumber} 
\def\beq{\begin{eqnarray}} 
\def\eeq{\end{eqnarray}} 
\def\ap{\left.} 
\def\at{\left(} 
\def\aq{\left[} 
\def\ag{\left\{} 
\def\cp{\right.} 
\def\ct{\right)} 
\def\cq{\right]} 
\def\cg{\right\}} 
\newtheorem{theorem}{Theorem} 
\newtheorem{lemma}{Lemma} 
\newtheorem{proposition}{Proposition} 
\def\R{{\hbox{{\rm I}\kern-.2em\hbox{\rm R}}}} 
\def\H{{\hbox{{\rm I}\kern-.2em\hbox{\rm H}}}} 
\def\N{{\hbox{{\rm I}\kern-.2em\hbox{\rm N}}}} 
\def\C{{\ \hbox{{\rm I}\kern-.6em\hbox{\bf C}}}} 
\def\Z{{\hbox{{\rm Z}\kern-.4em\hbox{\rm Z}}}} 
\def\ii{\infty} 
\def\X{\times\,} 
\newcommand{\bin}[2]{\left(\begin{matrix}{#1\cr #2\cr}
 \end{matrix}\right)} 
\newcommand{\fr}[2]{\mbox{$\frac{#1}{#2}$}} 
\def\Det{\mathop{\rm Det}\nolimits} 
\def\tr{\mathop{\rm tr}\nolimits} 
\def\Tr{\mathop{\rm Tr}\nolimits} 
\def\rot{\mathop{\rm rot}\nolimits} 
\def\PP{\mathop{\rm PP}\nolimits} 
\def\Res{\mathop{\rm Res}\nolimits} 
\def\res{\mathop{\rm res}\nolimits} 
\renewcommand{\Re}{\mathop{\rm Re}\nolimits} 
\renewcommand{\Im}{\mathop{\rm Im}\nolimits} 
\def\dir{/\kern-.7em D\,} 
\def\lap{\Delta\,} 
\def\arccosh{\mbox{arccosh}\:} 
\def\arcsinh{\mbox{arcsinh}\:} 
\def\arctanh{\mboz{arctanh}\:} 
\def\arccoth{\mbox{arccoth}\:} 

\def\d{${}^-\kern-.9em\partial\,$} 
\def\al{\alpha}
\def\be{\beta}
\def\ga{\gamma}
\def\de{\delta}
\def\ep{\varepsilon}
\def\ze{\zeta}
\def\io{\iota}
\def\ka{\kappa}
\def\la{\lambda}
\def\ro{\varrho}
\def\si{\sigma}
\def\om{\omega}
\def\ph{\varphi}
\def\th{\theta}
\def\te{\vartheta}
\def\up{\upsilon}
\def\Ga{\Gamma}
\def\De{\Delta}
\def\La{\Lambda}
\def\Si{\Sigma}
\def\Om{\Omega}
\def\Te{\Theta}
\def\Th{\Theta}
\def\Up{\Upsilon}

\title{Generalized modified gravity models: the stability issue}

\author{Guido Cognola\thanks{cognola@science.unitn.it}, 
Sergio Zerbini\thanks{zerbini@science.unitn.it}}
\address{Dipartimento di Fisica, Universit\`a di Trento \\ 
and Istituto Nazionale di Fisica Nucleare \\ 
Gruppo Collegato di Trento, Italia}

\begin{abstract}

A brief introduction on the issue of stability in generalized modified gravity 
is presented and the dynamical system methods are used in the investigation 
of the stability of spatially flat homogeneous
cosmologies within a large class of generalized modified gravity models 
in the presence of a relativistic matter-radiation fluid.

\end{abstract}


\section{Introduction}

To start with, we recall that recent cosmological data support the fact 
that there is a good evidence for a late accelerated expansion of the 
observable 
universe, apparently due to the presence of an effective positive and small 
cosmological constant of unknown origin. This is known as dark energy 
issue (see for example \cite{padmanabhan}). 
 
Among several existing explanation, the so called modified gravity models 
are possible realizations 
of dark energy (for a recent review and alternative approaches 
see \cite{CST,rev3,soti}), which may offer a quite natural 
geometrical approach again in the spirit of the original Einstein theory of
gravitation.
In fact, the main idea underlying these approaches to dark
energy puzzle is quite simple and consists in
adding to the gravitational Einstein-Hilbert action other gravitational terms 
which may dominate the cosmological evolution
during the very early or the very late universe epochs, 
but in such a way that General Relativity remains valid at intermediate epochs
and also at non cosmological scales.

The  $\La$-CDM model is  the simplest possibility
but, it is worth investigating   more  general modifications, 
possible motivations run from quantum corrections to string models. 
We shall first consider
the simpler modification of the kind $ F(R)=R+f(R)$, and then discuss the related generalizations. Models based on $F(R)$ are not new and they have been used in the past by many authors, for example as  models for inflation,  $f(R)=aR^2$ \cite{staro80}. Recently their interest in cosmology was triggered by the model  $f(R)=-\mu^4/R$, proposed in order  to describe the current acceleration of the observable universe \cite{capo03,turner}. 
For incomplete list of references, see \cite{R}. 

It is important to stress that  these $F(R)$ models  are conformally 
equivalent to Einstein's gravity, coupled with a self-interacting 
scalar field, Einstein frame formulation. We will consider only the Jordan frame, in which the dynamics of  gravity is described by $F(R)$ with minimally coupled matter. Observations are typically interpreted in this Jordan frame. 

We would like also to  mention the so called viable $F(R)$  models, which  have recently been proposed \cite{hu07,staro07,odin07,AB,so08,tret}, 
with the aim to describe the current acceleration with a  suitable choice of $F(R)=R+f(R)$, but also to be compatible  with local stringent gravitational tests of Einstein gravity $F(R)=R$. The main idea is the so called  disappearing of cosmological constant for low curvature, and mimicking the $\Lambda$-CDM model for high curvature. Thus, the requirements are:

a. $f(R) \to 0\,, \quad R \to 0\,$, compatibility with local tests;

b. $f(R) \to  -2 \Lambda_0\,, \quad R \to +\infty$, description of current acceleration;

c. Local stability of the matter.

As a illustration, we recall a recent example of viable model \cite{seba}
$$
f(R)=-\alpha \left( \tanh \left(\frac{b\left(R-R_0\right)}{2}\right) + \tanh \left(\frac{b R_0}{2}\right)\right)
$$
where $R_0$, and $\Lambda_0$  are suitable constants.
Its advantages are a better formulation in the Einstein frame and  a 
generalization that may also include the inflation era.

\section{The de Sitter stability issue for $F(R)$ models}
The stability of the de Sitter solution, relevant for Dark energy, 
 may be investigated in the  $F(R)$ models in several ways. 
We limit ourselves to the following  three approaches:

i. perturbation of the equations of motion  in the Jordan frame;

ii. one-loop gravity calculation around de Sitter background;

iii. dynamical system approach in FRW space-time. 

We shall briefly discuss   first two  approaches, and then concentrate to the third one, since it is the only that  can be easily extended to more general 
modified gravitational models, which is the main issue of this short review.

\subsection*{i. Stability  of $F(R)$ model in the Jordan frame} 

The starting point is the trace of the equations of motion, which is 
trivial in
Einstein gravity $R=-\kappa^2 T$, but, for a  general $F(R)$ model, reads
$$
3\nabla^2 f'(R)-2f(R)+Rf'(R)-R=\kappa^2 T\,.
$$
The new non trivial extra degree of freedom  is the scalaron: $1+f'(R)=e^{-\chi}$. 
Requiring $R=R_0=CST$, one has de Sitter existence condition in vacuum  
$$
R_0+2f(R_0)-R_0f'(R_0)=0\, .
$$
Perturbing around  dS:  $R=R_0+\delta R$, with
$\delta R = - \frac{1+f'(R_0)}{f''(R_0)} \delta\chi$, one arrives at
the scalaron perturbation equation
$$
\nabla^2 \delta\chi - M^2 \delta\chi = - \frac{\kappa^2}{6(1+f'(R_0))}T\, .
$$
One may read off the scalaron effective mass
$$
M^2\equiv \frac{1}{3}\left(\frac{1+f'(R_0)}{f''(R_0)} - R_0\right)\, .
$$
Thus, if  $M^2>0$, one has  stability of the dS solution and the related 
condition reads
$$
\frac{1+f'(R_0)}{R_0f''(R_0)}>1\, .
$$
If $M^2<0$,  there is a tachyon and  instability.
Furthermore, one may show that $M^2$ has to be very large in order to 
pass both the local and the astronomical tests and $1+f'(R)>0$, 
in order  to have a positive effective Newton constant. 
The same result has been obtained within a different more general perturbation approach in \cite{fara}.  

\subsection*{ii. One-loop $F(R)$ quantum gravity partition function}
Here we present the generalization to the modified gravitational case  
of the study of Fradkin and Tseytlin \cite{fra}, concerning Einstein gravity on dS 
space. One works in the Euclidean path integral formulation, 
with dS existence condition $2 F_0= R_0  F_0 $, assumed to be satisfied. 
The small fluctuations around this dS instanton may be written as
$$
g_{ij}= g_{0,ij}+h_{ij}\:,\hs
g^{ij}= g_0^{ij}-h^{ij}+h^{ik}h^j_k+{\cal O}(h^3)\:,\hs
h= g_0^{ij}h_{ij}\,.\nn 
$$
Making use of  the standard expansion of the tensor field $h_{ij}$ in irreducible components, and making an expansion up to second order in all the fields, 
one arrives at a very complicated Lagrangian density $ L_2 $, 
not reported here, describing Gaussian fluctuations around dS space. 
As usual, in order to quantise the model described by $ L_2$,  
one has to add gauge fixing and ghost contributions. 
Then, the computation of Euclidean one-loop partition function reduces 
to the computations of functional determinants. These functional determinants 
are divergent and may be  regularized by the well known zeta-function regularization. 
The evaluation requires a complicated  calculation  and,
neglecting the so called multiplicative anomaly, potentially present in zeta-function regularized determinants (see \cite{vanzo}), one arrives at the  one-loop effective action \cite{cogno05} (here written in the Landau gauge) 
\beq
\Ga_{on-shell}&=&\frac{24\pi F_0}{G R_0^2}
+ \frac12\,\log\det\ \aq \ell^2
\at-\lap_2+\frac{R_0}6\ct\cq \nn\\&&
      -\frac12\,\log\det\aq \ell^2 \at-\lap_1-\frac{ R_0}4\ct \cq
\nn\\&&
    +\frac12\,
\log\det \aq \ell^2 \at
-\lap_0-\frac{ R_0}3+\frac{2 F_0}{3 R_0 F_0''}\ct\cq\, \nn . 
\eeq
The last term is absent in the Einstein theory.
As a result,  in the scalar sector one has  an effective mass 
$M^2=\frac{1}{3}\at \frac{2  F_0}{ R_0  F_0''}- R_0 \ct$. 
Stability requires $M^2 >0$, in agreement with the previous scalaron analysis,
and with the inhomogeneous perturbation analysis \cite{fara}.

\section*{ The dynamical system approach}

The main idea of such a method 
is to convert the generalized Einstein-Friedman equations in an
autonomous system of first order differential equations and makes use of the 
theory of dynamical
systems (see \cite{gu,ellis,amendola,amendola1,dunsby,fay} and 
references therein). 
We remind that the stability or instability issue is really 
relevant in cosmology. 
For example, in the $\Lambda$CDM model it ensures that no future 
singularities will be present in the solution.
Within cosmological models, the stability or instability around a 
solution is of interest at early and also at late times. 

The stability of de Sitter solutions has been 
discussed in several places, an incomplete list being 
\cite{fara,cognola,NO,capo,NO1,NO2,bazeia07,rador,soko,b}. 
More complicated is the analysis of other critical points, associated with the presence of non vanishing matter and radiation. For example, in the paper 
\cite{sami}, a non local model of modified gravity  $F(R)$ has been investigated by means of this approach. 

Here we shall extend to general modified gravity in the presence of matter
the method given in Ref.~\cite{amendola}, which permits
to determine all critical points of a $F(R)$ model.
Ordinary matter is important in reconstructing the expansion 
history of the Universe and probing the phenomenological relevance of the 
models (see for example the recent papers 
\cite{amendola,amendola1,dunsby,fay}, where the $F(R)$ case has been 
discussed in detail). 
Our generalisation consists in the extension of that method 
in order to include all 
possible geometrical invariants. This means that $F$ 
could be a generic scalar function of curvature, Ricci and 
Riemann tensors. 

There are some theoretical (quantum effects and string-inspired) 
motivations in order to investigate 
gravitational models depending on higher-order invariants.
The ``string-inspired'' scalar-Gauss-Bonnet gravity case $F(R,G)$ 
has been suggested in Ref.~\cite{sasaki} as a model for
gravitational dark energy, while some time ago it
has been proposed as a possible solution of the initial singularity 
problem \cite{ART}. 
The investigation of different
regimes of cosmic acceleration in such gravity models
has been carried out in 
Refs.~\cite{sasaki,fGB,sami1,Mota,Calcagni,Neupane,GB,cognola06,cognola066,sami2,E}.
In particular, in \cite{cognola06} a first attempt to the study of 
the stability of such kind of models has been carried out
using an approach based on quantum field theory. 
 
The method we shall use in the present paper is based on a 
classical Lagrangian formalism, see, for example, 
\cite{vilenkin85,capozziello02,cognolaV}, inspired by the paper 
\cite{staro80}, where quantum gravitational effects were considered for the 
first time.
With regard to this, it is well known that one-loop and two-loops quantum 
effects induce higher derivative gravitational terms in the effective 
gravitational Lagrangian. Instability due to 
quadratic terms have been investigated in \cite{staro1}. 
A particular case has been recently studied in \cite{aco} 
and general models depending on quadratic invariants
have been investigated in \cite{herv,topo}. 

A stability analysis of nontrivial vacua in a general class of 
higher-derivative theories of 
gravitation has already been presented in \cite{waldram}. 
Our approach is different from the one presented there 
since we are dealing with scalar quantities and moreover it is 
more general, since it is not restricted to the vacuum invariant submanifold.

Finally, it should be stressed that the stability studied here is the one
with respect to homogeneous perturbations. For the $F(R)$ case, the stability
criterion for homogeneous perturbations is equivalent to the inhomogeneous one
\cite{fara}. In the following we will summarize the results obtained in 
\cite{cogno08}.
\subsection{The dynamical system approach: The general case}
 To start with, let us consider a Lagrangian density which is an 
arbitrary function of all algebraic invariants built up with the Riemann 
tensor of the FRW space-time we are dealing with, that is
\beq 
{\cal L}=-\frac{1}{2\chi}\,F(R,P,Q,...)+{\cal L}_m\,,
 \eeq
where $R$ is the scalar curvature, $P=R^{\mu\nu}R_{\mu\nu}$ and 
$Q=R^{\alpha\beta\gamma\delta}R_{\alpha\beta\gamma\delta}$ are the two 
quadratic invariants 
and the dots means other independent algebraic invariants of higher order, and
${\cal L_m}$ is the matter Lagrangian which depends on $\rho$ and $p=p(rho)$,
the density and pressure of the matter. 

For the sake of convenience we write the metric in the form
\beq 
ds^2=-e^{2n(t)}dt^2+e^{2\al(t)}d{\vec x\,}^2\,,
\hs N(t)=e^{n(t)}\,,\qquad a(t)=e^{\al(t)}\,.
 \eeq 
In this way $\dot\al(t)=H(t)$ is the Hubble parameter and
a generic invariant geometrical quantity $U$ has the form
\beq 
U=e^{-2pn(t)}\,u(\dot n,\dot\al,\ddot\al)=e^{-2pn(t)}\,u(\dot n,H,\dot H)
=H^{2p}\,e^{-2pn(t)}\,u(X)\,,
\label{INV}\eeq
where $2p$ is the dimension (in mass) of the invariant under consideration
and $X=(\dot H/H^2-\dot n/H)$ (see \cite{cogno08}).
In particular one has
\beq\begin{array}{l}
R=6H^2e^{-2n}(2+X)\,,
\\ 
P=12H^4e^{-4n}(3+3X+X^2)\,,
\\ 
Q=12H^4e^{-4n}(2+2X+X^2)\,,
\\ 
\end{array}\label{INV-Q}\eeq
Using this notation, the action reads
\beq 
S=-\int\,d^3x\int\,dt\, L(n,\dot n,\al,\dot\al,\ddot\al)
 =\frac{1}{2\chi}\,\int\,d^3x\int\,dt\,e^{n+3\al}\,F(n,\dot n,\dot\al,\ddot\al)
+S_m \,, 
\label{GrA}\eeq
and the Lagrange equations corresponding to the two Lagrangian variables 
$n(t)$ and $\al(t)$ are given by
\beq 
E_n=\frac{\partial L}{\partial n}
 -\frac{d}{dt}\,\frac{\partial L}{\partial\dot n}
 =2\sqrt{-g}\,T_{00}g^{00}=2\rho\,\sqrt{-g}
\,,
\label{EnM}\eeq
\beq 
E_{\al}=\frac{\partial L}{\partial \al}
 -\frac{d}{dt}\,\frac{\partial L}{\partial\dot\al}
 +\frac{d^2}{dt^2}\,\frac{\partial L}{\partial\ddot\al}
 =\sqrt{-g}\,T_{ab}g^{ab}=-6p\,\sqrt{-g}\,.
\label{EaM}\eeq
It has to be noted that since $n(t)$ is a ``gauge function'', 
which corresponds to the
choice of the evolution parameter,
Eqs.~(\ref{EnM}) and (\ref{EaM}) are not independent
and in fact they satisfy the differential equation
\beq
\frac{dE_n}{dt}=\dot nE_n+\dot\alpha E_n\,,
\eeq
which is equivalent to the conservation of energy-momentum tensor.
Furthermore, we may use the gauge freedom and fix the cosmological 
time by means of the condition $N(t)=1$, that is $n(t)=0$. 
From now on it is understood that all quantities will be evaluated in 
such a gauge and so the parameter $t$ corresponds to the standard cosmological time. 
In this way Eqs.~(\ref{EnM}) and (\ref{EaM}) read
\beq 
H\dot F_{\dot H}-HF_H+F-\dot HF_{\dot H}+3H^2F_{\dot H}=2\rho\,,
\label{EnM2}\eeq
\beq 
\ddot F_{\dot H}-\dot F_H+6H\dot F_{\dot H}-3HF_H
+3F+3\dot HF_{\dot H}+9H^2F_{\dot H}=-6p\,.
\label{EaM2}\eeq
The latter equations are the generalisation to arbitrary action of
the well known Friedmann equations.

Now we shall replace Eqs.~(\ref{EnM2})-(\ref{EaM2}) with an autonomous system
of first order differential equations.
To this aim we first observe that in
pure Einstein gravity, that is for $F=R$,
(\ref{EnM2})-(\ref{EaM2}) read
\beq
H^2F_{\dot H}=F_X=2\rho\segue \Om_\rho=1\,,
\label{EnS}\eeq
\beq
(3H^2+2\dot H)F_{\dot H}=(3+2X)F_X=-6p
\segue \Om_p=-1-\frac23\,X\,,
\label{EaS}\eeq
where we have introduced the dimensionless variables
\beq
\Om_\rho=\frac{2\rho}{H^2F_{\dot H}}=\frac{2\rho}{F_X}\,,\hs\hs
\Om_p=\frac{2p}{H^2F_{\dot H}}=\frac{2p}{F_X}\,,
\label{Om-rp}\eeq
which in this special case are given by the usual values
$\Om_\rho=\rho/3H^2$ and $\Om_p=p/3H^2$.
From Eqs.~(\ref{EnS}) and (\ref{EaS}) it follows
\beq
w\equiv\frac{p}{\rho}=\frac{\Om_\rho}{\Om_p}=-1-\frac23\,X\,.
\label{w}\eeq
In the general case, Eqs.~(\ref{EnM2}) and (\ref{EaM2}) have
more terms with respect to (\ref{EnS}) and (\ref{EaS})
and it is quite natural to interpret them as corrections
due to the presence of higher-order terms in the action.
Then we define
\beq
\Om_{\rho}^{eff}=\Om_\rho+\Om_\rho^{curv}=1\,,\hs\hs
\Om_p^{eff}=\Om_p+\Om_p^{curv}=
-1-\frac23\,X\,,
\eeq
\beq
w_{eff}\equiv\frac{\Om_p^{eff}}{\Om_\rho^{eff}}=-1-\frac23\,X\,,
\label{weff}\eeq
where $\Om_\rho^{curv}$ and $\Om_p^{curv}$ are complicated expressions,
which only depend on the function $F$. They can be derived
from (\ref{EnM2}) and (\ref{EaM2}),
but they explicit form is not necessary for our aims.
The effective quantity $w_{eff}$ is equal to the ratio between the
effective density and the effective pressure and it could be
negative even if one considers only ordinary matter.
It is known that the current-measured value of $w_{eff}$
is near $-1$.

In order to get all critical points of the system now we follow the
method, as  described, for example, in \cite{amendola}. First of all, 
we introduce the
dimensionless variables
\beq
\Om_\rho=\frac{2\rho}{H^2F_{\dot H}}=\frac{2\rho}{F_X}\,,\hs\hs
\Om_p=\frac{2p}{H^2F_{\dot H}}=\frac{2p}{F_X}\,,
 \eeq
\beq
X=\frac{\dot H}{H^2}\,,\hs
Y=\frac{F-HF_H}{H^2F_{\dot H}}=\frac{F}{F_X}-X\,,\hs
Z=\frac{\dot F_{\dot H}-F_H}{HF_{\dot H}}=\frac{F'_X}{F_X}-2X-\xi\,,
\label{XYZ}\eeq
where the prime means derivative with respect to $\al$ and the quantity
\beq 
\xi=\xi(X,Y)=\frac{F_H}{HF_{\dot H}}=\frac{HF_H}{F_X}\,,
\label{xi}\eeq
has to be considered as a function of the variables $X$ and $Y$. 
In general it is a function of $X$ and $H$, but this latter 
quantity can be expressed in terms of $X$ and $Y$ as a direct 
consequence of the definition of $Y$ itself.
Then we derive an autonomous system by taking the 
derivatives of such variables.
From Eq.~(\ref{EnM2}) we have the constraint
\beq 
\Om_\rho=Y+Z+3\segue\Om_{curv}=-(Y+Z+2)\,,
\label{vincolo}\eeq
which reduces to the standard one when $F$ is linear in $R$ (general relativity with
cosmological constant). 

Deriving the variables above by taking into account of 
(\ref{EaM2}) and (\ref{vincolo}) we get the system
of first order differential equations 
\beq\ag\begin{array}{l} 
X'=-2X^2-\ga X+\be(Z+\xi)
\\
Y'=-(2X+Z+\xi)Y-XZ
\\
Z'=-3(1+w)(Z+Y+3)-(Z+\xi)(Z+3)-X(Z+6)
\end{array}\cp\label{AutS}\eeq
where $X'\equiv\frac{dX}{d\al}=\frac{1}{H}\frac{dX}{dt}$ (and so on) 
and for simplicity we have set $p=w\rho$,
with constant $w$. For ordinary matter $0\leq w\leq1/3$
($w=0$ corresponds to dust, while $w=1/3$ to pure radiation),
but in principle one could also consider ``exotic'' matter with $w<0$
or cosmological constant which corresponds to $w=-1$. 
We have also set
\beq 
\be=\be(X,Y)=\frac{F_{\dot H}}{H^2F_{\dot H\dot H}}
 =\frac{F_X}{F_{XX}}\,,\hs\hs
\ga=\ga(X,Y)=\frac{F_{H\dot H}}{HF_{\dot H\dot H}}
 =\frac{HF_{HX}}{F_{XX}}=\be\xi_X+\xi\,.
\label{bega}\eeq
It is understood that $F_{\dot H\dot H}\neq0$ has been assumed.
The quantity $\Om_\rho$ at the critical points will be determined by means 
of Eq.~(\ref{vincolo}).

The critical points are obtained by putting 
$X'=0,Y'=0,Z'=0$ in the system (\ref{AutS}). 
The number and the position of such points depends on the Lagrangian throughout
the functions $\be,\ga$ and $\xi$. 
In principle, given $F$ one can derive 
all critical points, but in practice for a generic 
$F$ the algebraic system could 
be very complicated and the solutions quite involved.
We shall consider in detail some particular cases at the end of the Section.

As already said above, the critical points of (\ref{AutS}) 
depends on $\xi,\be,\ga$, which in general are complicated 
functions of $X$ and $Y$, then it is not possible to determine
general solutions without to choose the model, nevertheless it is
convenient to distinguish two distinct classes of solutions
characterised by the values of $w\neq-1$ and $w=-1$.
For the sake of completeness we consider $w\leq1/3$ and so 
we write the solutions also for ``exotic'' matter, that is
quintessence ($-1<w<0$) and phantom ($w<-1$). 
Of course, such solutions have to be dropped if
one is only interested in ordinary matter/radiation.
We have
\begin{itemize}
\item $w\neq-1$ --- The critical points
are the solutions of the system of three equations
\beq 
\ag\begin{array}{l}
2X^2+\ga X-\be(Z+\xi)=0
\\
(2X+Z+\xi)Y+XZ=0
\\
3(1+w)(Z+Y+3)+(Z+\xi)(Z+3)+X(Z+6)=0
\end{array}\cp\hs\hs w\neq-1\,,
\label{C1}\eeq
where $\xi,\be,\ga$ are functions of $X,Y$ determined by Eqs.~(\ref{xi})
and (\ref{bega}). 
The stability matrix has three eigenvalues and the point is 
stable if the real parts of all of them are negative. 

The latter system has always the de Sitter
solution $P_0\equiv(X=0,Y=1,Z=-4)$, where $\Om_\rho=0$
and $w_{eff}=-1$. 
Note however that such a solution could exist also in the presence
of matter, since the existence of $P_0$ critical point 
only implies that the critical value for $\Om_\rho$ vanishes.

\item $\Om_\rho\neq0$, $w=-1$ --- 
The critical points are given by
\beq 
\ag\begin{array}{l}
2X^2+\ga X-\be(Z+\xi)=0
\\
(2X+Z+\xi)Y+XZ=0
\\
(Z+\xi)(Z+3)+X(Z+6)=0
\end{array}\cp
\hs\hs w=-1\,.
\label{C-1}\eeq
For this class of solutions, the non-singular stability matrix
has three eigenvalues and the point is stable if the real parts of all of
them are negative.

We see that there is at least one singular case 
(critical line) when $X=0$ and $Z=-\xi=-4$. 
In fact in such a case $Y$ or $\Om_\rho$ are undetermined since
\beq
\Om_\rho=Y+3-\xi(0,Y)=Y-1\segue Y=1+\Om_\rho\,,
\hs\hs\Om_\rho\mbox{ arbitrary}\,.
\label{special}\eeq
Such a solution can be seen as a generalisation of 
the de Sitter solution for a model with cosmological constant.
The de Sitter critical point for the model $\tilde F=F-2\La$
reads $(X=0,\tilde Y=1,\tilde Z=-4)$.
Such a solution follows from Eq.~(\ref{special}) if we choose
$\rho_0\equiv\La$. In fact,
on the critical point $(X=0,Y=1+\Om_\rho,Z=-4)$ (Eq.~(\ref{special})) 
and from definitions (\ref{XYZ}) we get
\beq 
\Om_\rho=\frac{2\rho}{H^2F_{\dot H}}=\frac{F}{H^2F_{\dot H}}-1
\segue\tilde Y\equiv\frac{\tilde F}{H^2\tilde F_{\dot H}}=1\,,
 \eeq
which corresponds to de Sitter critical point for $\tilde F$.
Eq.~(\ref{special}) is more general than the
case with pure cosmological constant since $\rho$ is not
necessary a constant, and for this special class of solutions $w_{eff}=-1$. 
Note also that the stability matrix has always a vanishing eigenvalue
and the stability of the system is determined by the
other two eigenvalues.

For some models, but just for technical reasons, 
it could be convenient to treat the cosmological constant as matter,
using the previous identification we have done. 

\end{itemize}

\subsubsection*{Explicit examples}

In order to see how the method works, 
now we give explicit solutions for some models and, when possible we also 
study the stability of the critical points. 
We restrict our analysis to the values $0\leq w\leq1/3$ and
to the special value $w=-1$, which corresponds to the pure cosmological constant,
but in principle any negative value of $w$ could be considered,
even if this will be in contrast with the aim of modified gravity.
In fact, modified gravity can generate an effective negative value of $w$ 
without the use of phantom or quintessence.

It as to be stressed that in general, due to technical difficulties,
one has to study the models by a numerical analysis. 
Only for some special cases one is able to
find analytical results. 
Here we report the results
for some models of the latter class in which the analytical 
analysis can be completely carried out. We also study
more complicated models and for those we limit
our analysis to the de Sitter solutions. 

In the following we shall use the compact notation
\beq 
P\equiv(X,Y,Z,\Om_\rho,w_{eff})\,,\hs
P_0\equiv(0,1,-4,0,-1)\,,\hs
P_\La=(0,1+\Om_\La,-4,\Om_\La,-1)\,.
 \eeq
The latter is an additional critical point
that we have for the choice $w=-1$ and can be seen as the de Sitter solution
in the presence of cosmological constant.

\begin{description}

\item
$F=R-\mu^4/R$ --- This is the well known model introduced in 
\cite{capo03,turner} and discussed in \cite{amendola}. 
For this model the system (\ref{AutS}) with arbitrary $w$ 
has six different solutions, 
but only two of them effectively correspond to 
physical critical points, if $0\leq w\leq1/3$. 
In principle there are other critical points for negative
values of $w$ (phantom or quintessence) and 
moreover there is also a particular solution for $w=-1$ which 
corresponds to the model with a cosmological constant $\La$. 
 
Solving the autonomous system one finds
\begin{itemize}
\item 
$P=P_0$: unstable de Sitter critical point.
The critical value for the scalar curvature reads $R_0=\sqrt3\,\mu^2$.
\item 
$P=(-1/2,-1,-2,0,-2/3)$: stable 
critical point. At the critical value, $H_0=0$. 
\item 
$P=(3(1+w)/2,-(5+3w),-2(5+3w),-3(4+3w),-(2+w))$: 
unstable critical point where $H_0=0$. 
\item 
$P=P_\La$: unstable critical point. At the critical value one has 
$
H_0^2=(\La/6)(1+\sqrt{1+3\mu^4/4\La^2}\,.
$
\end{itemize}

\item
$F=R+aR^2+bP+cQ$ --- (Starobinsky-like model).
Here we have to assume $3a+b+c\neq0$ otherwise the quadratic term
becomes proportional to the Gauss-Bonnet invariant. 
For $0\leq w\leq1/3$, this model has only one critical point. 
In order to have a de Sitter solution, we have to introduce a 
cosmological constant $\La$. We have in fact
\begin{itemize}
\item 
$P=P_0$: Minkowskian solution 
with $R_0=0$, which is stable if $3a+b+c>0$.
\item 
$P=P_\La$:
de Sitter critical point with $R_0=6\La$, which 
is stable if $3a+b+c>0$, in agreement with \cite{topo}.
\end{itemize}

\item
$F=R-d^2Q_3$ --- This is the simplest toy model with 
the cubic invariant
$Q_3=R^{\al\be\ga\de}R_{\al\be\mu\nu}R^{\mu\nu}_{\ga\de}$.
For this model we have the following critical points:
\begin{itemize}
\item
$P=P_0$: unstable de Sitter solution with $R_0=6/d$.
\item
$P=P_0$: stable Minkowskian solution with $R_0=0$.
\item
$P=(0.05,0.60,-3.60,0,-1.03)$: stable solution with $H_0=0$.
\item
$P=P_\La$: this point exists, but not for any value of $d$ and $\La$.
Also the value of $H_0$ and the stability depend on the parameters.
 \end{itemize}

\item
$F=R+aR^2+bP+cQ-d^2Q_3$ --- This is a generalisation of the previous
two models. It may be motivated by the two-loop corrections in quantum gravity 
\cite{sagno,van}. The de Sitter critical points have been studied in 
Ref.~\cite{monica}.
The algebraic equations (\ref{AutS}) are too 
complicated to be solved analytically, but it is easy
to verify that there are at least the following solutions:
\begin{itemize}
\item
$P=P_0$: de Sitter solution with $R_0=6/d$. This is stable 
if $3a+b+c+3d>0$.
\item 
$P=P_0$: Minkowskian solution 
with $R_0=0$, which is stable if $3a+b+c>0$.
\item
$P=P_\La$: also in this case this point exists and 
is stable depending on the parameters (see \cite{monica}).

 \end{itemize}

\end{description} 

\s{Conclusion}
In this contribution, after a short review on the $F(R)$ case, we 
have presented the dynamical system approach which has permitted 
to arrive at a first order autonomous system of differential equations 
classically equivalent to the equations of motion for models of 
generalized modified gravity based on an arbitrary function $F(R,P,Q,Q_3...)$, 
namely built up with all possible geometric invariant quantities of 
the FRW space-time. We have shown that, in the special case
of $F(R)$ theories, the method gives rise 
to the well known results \cite{amendola,amendola1,dunsby}, 
but in principle it can be applied to the study of 
much more general cases. 

As illustrations, we have discussed some
simple models, for which a complete analytical 
analysis concerning the critical points has been carried out. 
However, in general,
due to technical difficulties, a numerical analysis 
is required. Among the models investigated, we would like to remind that 
we were able to deal with one which involves a cubic invariant in the 
curvature tensor and, to our knowledge, this has never been considered before,
and this shows the power of the present approach.

\section*{Acknowledgments}
This review is based mainly on results obtained in collaboration with 
E. Elizalde, S. Nojiri, and S. D. Odintsov and we would like to thank them.

\end{document}